\documentclass{vgtc}                          

\ifpdf
  \pdfoutput=1\relax                   
  \usepackage{graphicx}                
  \DeclareGraphicsExtensions{.pdf,.png,.jpg,.jpeg} 
\else
  \usepackage{graphicx}                
  \DeclareGraphicsExtensions{.eps}     
\fi%

\graphicspath{{figures/}{pictures/}{images/}{./}} 

\usepackage{comment}
\usepackage{wrapfig}
\usepackage{microtype}                 
\PassOptionsToPackage{warn}{textcomp}  
\usepackage{textcomp}                  
\usepackage{mathptmx}                  
\usepackage{times}                     
\usepackage{cite}                      
\usepackage{tabu}                      
\usepackage{booktabs} 

\usepackage{enumitem}
\setlist[itemize]{noitemsep, topsep=0pt}

\onlineid{0}

\vgtccategory{Research}

\vgtcinsertpkg

\title{Can AI Mitigate Human Perceptual Biases?  A Pilot Study}

\author{
Ross Geuy
\thanks{e-mail: geuy.8@buckeyemail.osu.edu} 
\\
\scriptsize The Ohio State University%
\and Nate Rising\thanks{e-mail: rising.14@buckeyemail.osu.edu}\\ %
\scriptsize Carnegie Mellon University%
\and Tiancheng Shi \thanks{e-mail: ts3474@columbia.edu}\\ %
\scriptsize Columbia University 
\and Meng Ling \thanks{e-mail: ling.253@buckeyemail.osu.edu}\\ %
\scriptsize The Ohio State University and Shell plc 
\and Jian Chen\thanks{e-mail: chen.8028@buckeyemail.osu.edu} \\
\scriptsize The Ohio State University%
}

\teaser{
  \centering
  \vspace{-15pt}
\includegraphics[width=0.7\linewidth]{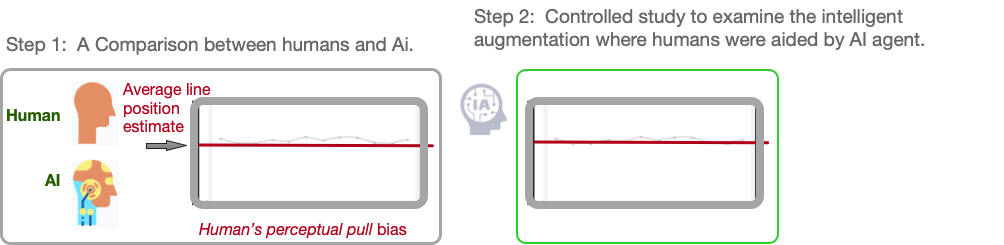}
\vspace{-10pt}
  \caption{People tend to underestimate the average line positions (toward the bottom of the display)\cite{xiong2019biased}. In this pilot study, we examined if AI decisions can help debias innate human perceptual biases when people perform the same task with AI aids.  We first compared the prediction accuracy between AI and humans performing the same task (Step 1). We then added a one-off visualization (Step 2) to the environment and asked people to estimate with AI's assistance.}
  \label{fig:teaser}
}

\abstract{
{We present results from a pilot experiment to measure if machine recommendations can debias human perceptual biases in visualization tasks.
We specifically studied the ``pull-down'' effect, i.e., people underestimate the average position of lines~\cite{xiong2019biased}, for the task of estimating the ensemble average of data points in line charts. These line charts can show for example temperature or precipitation in 12 months. Six participants estimated ensemble averages with or without an AI assistant. The assistant, when available, responded at three different speeds to assemble the conditions of a human collaborator who may delay his or her responses.} 
Our pilot study showed that participants were faster with AI assistance in ensemble tasks, compared to
the baseline without AI assistance.
Although ``pull-down'' biases were reduced, the effect of AI assistance was not statistically significant. Also, delaying AI responses had no significant impact on human decision accuracy. We discuss the implications of these preliminary results for subsequent studies.
} 

\CCScatlist{
  \CCScatTwelve{Human-centered computing}{Visu\-al\-iza\-tion}{Visu\-al\-iza\-tion techniques}{evaluation methods};
  \CCScatTwelve{Human-AI teaming}{Pilot study}{charts}{biases};
}

\begin{document}


\firstsection{Introduction}

\maketitle
We envision a condition where
people work ``with AI'' to correct human innate decision biases when reading visualizations. 
The rationale is that machines running deep neural network algorithms (DNNs) are surprisingly effective at a wide variety of tasks traditionally associated with human visual intelligence. For example, DNNs achieved superhuman level accuracy while reading charts or counting communities from node-link diagrams~\cite{jiang2023Sampling}.
If these algorithms can be successful in these visual perception tasks, \textit{can they help humans overcome innate perceptual biases reported in recent studies?} 
Also, humans' graphical perception differ by cognitive accuracy and speed toward decoding visualizations and layout~\cite{cleveland1984graphical}. \textit{Would AI-in-the-human-loop
improve human consistency?}


To answer these questions, we attempted to understand AI expertise and the dependence of humans on AI judgement.
Studies show that people can have different reactions to visual images based on the extent of their background information~\cite{hardstone2021long, xiong2019curse}. Specifically, experts who spend significant amounts of time with a particular dataset have been shown to gloss over important low-level details in their visualizations or reports.
In visualization and vision science, perceiving averages or other statistical features from a group of similar items, called \textit{ensemble perception}~\cite{chen2019measuring, whitney2018ensemble}, is a robust visual phenomenon that operates across a host of visual dimensions. One such case is to estimate the ensemble average of the data points along the line charts~\cite{correll2012comparing}. However, this perception is \textit{ pulled down} in the direction of the ensemble mean when estimating partial data~\cite{xiong2019biased}. That is, the viewer's estimate of individual points or a data series is towards the center of the figure containing a single data series or towards another data series in a multivariate graph.
Human visual system is sensitive to the central tendency of the display. 
Given their practical significance, we chose to study the perceptual biases in line charts.


In real-world uses, for example, perceiving temperature increases due to climate changes or drug abuse in Ohio after FDA policy regulations, we would have taken two steps in the evaluation~\cite{lakkaraju2022rethinking}.
The first step (Figure~\ref{fig:teaser}) is
the comparison between decision makers to ask if the prediction of an algorithm is significantly better than that of humans. In fact, we found in the bar chart conditions that on average machines’ decisions were better and may not conform to human predictions. The machine accuracy dropped dramatically when the test data fell out of the training data range or when the machines were asked to observe charts with smaller bar heights~\cite{jiang2023Sampling}. 

The second step is to take into account potential confounding factors, given that
humans can have a more complicated set of personal biases over an input than AI models. For example, judges may have their own opinions in court cases~\cite{slack2021counterfactual}. In economics, such factors or variables that are not observed in studies but are relied on by people in their decision-makings are called \textit{private information}~\cite{tversky1974judgment}. Human decision makers depend on many variables that are not recorded, manifesting itself as an information-impaired version of themselves in an empirical study. Since human characteristics related to their \textit{private information} are largely unknown, many visualization studies, including our own, have used methods to take away people's private information (e.g., background knowledge) using simulated data to replace real-world data, when evaluating visual encoding and display conditions. To collect data from diverse populations, we took into account the trade-off between validity and domain specificity by guiding viewers with stories of climate data and using simulated temperature records. 


\textbf{Contributions. } Our first pilot study has begun to systematically explore how AI assistance can help mitigate human biases. Our research contributes to the following: 

\begin{itemize}
\item 
A new design that provides people access to AI's actionable insights for unbiased human decision.
\item 
Statistical and anecdotal evidence, including time, accuracy, and viewers' attitudes that will be useful in designing future visualization of AI results for more accurate graphical perception.
\item 
A first look at the broader issues of using intelligent augmentation to assist people with elementary tasks. 
\end{itemize}


\begin{figure*}[!ht]
 \centering 
 \includegraphics[width=0.98\textwidth]{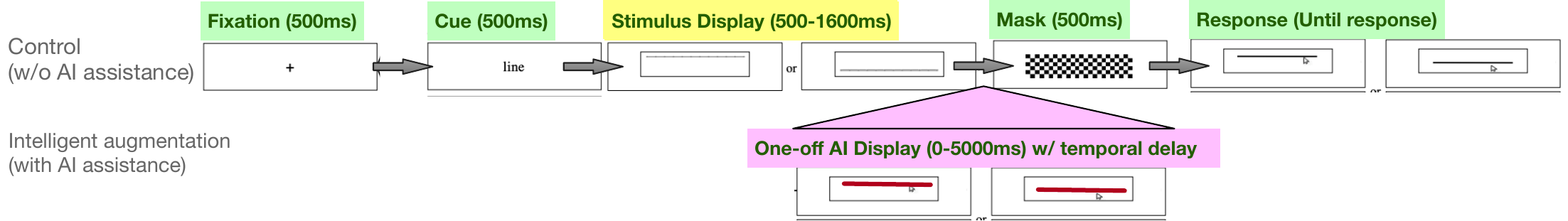}
 \vspace{-10pt}
 \caption{Evaluation procedure and design for our pilot study. The green display boxes followed the same procedure as Xiong et al.~\cite{xiong2019biased}. A random temporal delay was added to AI responses. AI assistant is shown in pink.}
 \label{fig:exp}
\end{figure*}

\section{Related Work}

We summarize the literature to help arrive at factors that influence human decisions when they collaborate with another agent, where an agent can be a person or an algorithm.

\subsection{Human perceptual biases}
People are perceptually biased~\cite{xiong2019biased, xiong2019curse}.
Xiong et al.~\cite{xiong2019biased} asked respondents to estimate the average value of a series of line graphs, bar graphs, and a combination of the two on the same graph. They found that humans tended to underestimate the true mean of line graphs. Our goal is to help better understand the circumstances that create cases and deploy tools 
in the most effective way to avoid those human biases.

\subsection{Human-AI Comparisons}
Haehn et al.~\cite{haehn2018evaluating} pioneered human-AI comparison in visualizations and reported that machines outperformed humans.
Post-hoc explanation of black-box AI methods is useful for studying human-AI differences. This line of work is a precursor to our work for understanding what makes an AI model useful in the context of specific tasks. It includes (1) predicting the AI output, (ii) verifying an algorithm's output consistency with the visual explanation, and (iii) determining model robustness to input: if and how the output would change if we change the input.  
Hullman et al.~\cite{hullman2022worst} also encourage the use of social science methods to evaluate AIs.

\subsection{AI Assistance for Human Decision Making}

More closely related to this study, there has been significant work exploring human responses to AI decisions and AI assistance. 
Levy et al.~\cite{levy2021assessing} showed that systems where humans actively interacted with an AI assistant outperformed AI systems and human systems alone. However,
interacting with an AI assistant over an extended period led the human to simply agree with whatever the AI suggests. That is, they trust AI so much that they will blindly trust whatever the AI prediction is. 
Ferreira and Monteiro~\cite{ferreira2021human}
discuss the responsibility of AI-analysis in terms of justification and explanation. Since human decision makers often must justify their decisions, they should also justify AI's decision if AI contributes to the decisions they make. Explanation allows the user to understand the intrinsic processes of AI, while justifications are extrinsic sources of information to validate AI results. 

Humans' trust in AI differs from their trust in humans. Humans tend to have lower trust in results recommended by AI. This effect also depends on expertise - expert viewers rate the usefulness of AI much lower than novice viewers~\cite{gaube2021ai}. 
Strategies have been developed to make people think more critically. For example, 
Fridman et al.~\cite{fridman2019arguing} in their `Arguing Machines' trained two cases of competing AI agents and used disagreement between the two as a signal to seek the viewer's supervision. The overall goal was to apply an ‘arguing machines’ model to evaluate cases where a human-decision was sought. The overall system error was greatly reduced.
While their models provided text and label suggestions, they are not sufficient to prevent humans from accepting an erroneous result, regardless of the model's confidence. Other learning-theory-based metrics have also been developed. 
Burns et al.~\cite{burns2020evaluate} leveraged the philosophical aspect of evaluating data visualizations across different levels of understanding. In the study, the level of understanding the audience gained after seeing the visualization, as evaluated by Bloom's Taxonomy, were compared to the ``true'' meaning that the data analyst originally wished to convey. 

\subsection{Influence from Collaborators}

One fascinating finding on collaboration among decision makers is that the collaboration process can reveal private information.
Frydman et al.~\cite{frydman2022using} suggested that response time is inherent to infer a viewer's own decision. They investigate how the speed at which people make decisions influence the decision-making process of others, i.e. ‘group-think’. At its core, this study can be viewed as a classification problem; given their own private information and previous participant’s decisions, each person chooses to either agree with their private information or disagree with it. The study found that when the group’s decision is contradictory to an individual’s private opinion, there is an increase in their response time as they consider to ultimately keep their decision or ‘side with the herd.’ 
This is used in psychology to reveal ``\textit{private information}'' of a decision maker among a group of human participants. Their goal was not to use machine intelligence to aid people. 

\subsection{Summary}
This line of pioneering work is mainly studied in conditions where human collaborators work together in decision-making, or in medical-imaging and so on. Our work would be among the first to study debiasing human biases with AI assistance in visual tasks.

\section{Method: Show me the Insight, Not Just Data}

This section describes our method to make machine useful to human viewers:
our creative statement is to make the visual interface show not only the encoded data but also the actionable insights. Our proposed study framework was to measure human performance with and without machine assistance. 
In this fashion, viewers in complex scenarios can indulge in better decision-making. We are interested in viewers' behavior changes activated by machine's recommendations.

\subsection{Hypotheses}
To establish a baseline for human perception in conditions in which AI is added to the human decision loop, we compare results with and without AI assistance. 
We had the following hypotheses. 
\begin{itemize}
\item
\textbf{H1.} 
AI will improve the performance of human participants in terms of accuracy and response time.
\item
\textbf{H2.}
The ``pull-down'' effect from Xiong et al.~\cite{xiong2019biased} exists in the without-AI conditions. 
\item
\textbf{H3.} 
The ``pull-down'' effect will be diminished in the with-AI conditions.
\item 
\textbf{H4.} Humans will be willing to ``\textit{herd}'' an AI.
\end{itemize}

In these hypotheses (H1, H3, and H4), we assumed that AI inferences were so accurate that humans' making use of it would have the most accurate predictions. H2 

\subsection{Visualization Design with AI Insights}
Test images were generated as square line charts composed of 12 data points having a resolution of 100 by 100 pixels (see below). 
\setlength{\columnsep}{10pt}%
\begin{wrapfigure}{r}{0.25\columnwidth}
    \vspace{-10pt}
    \includegraphics[width=0.25\columnwidth]{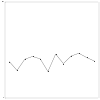}
    \vspace{-20pt}
  \label{fig:sampleImage}
\end{wrapfigure}
The 12 data points in each line chart were randomly generated using the method in Xiong et al.~\cite{xiong2019biased}. 
To control the vertical position of the line chart (i.e., high, middle, or low on the chart), a randomized ``position value'' was added to the Y coordinates of the 12 data points with the constraint that they stay within bounds. 

Our visualization supports showing the AI inference results. The visualization is analogous to having the AI insights available to the viewers' task at hand. In particular, we are interested in making the insight directly accessible via direction visualization. A red line is overlaid on the line chart to show the AI ensemble average reading result.

\subsection{Empirical Study Design and Procedure} 

Our experimental setup is shown in Figure~\ref{fig:exp} with two dependent variables: 1) The presence or absence of AI assistance and 2) the temporal delay of AI assistance. Participants viewed 50 randomly generated test images and were asked to provide their mean estimate. The first 15 were the control tests without AI assistance and the remaining 35 tests were provided with AI assistance and were shown in a random order within the same group.

\paragraph{W/O-AI Assistant}
The Without-AI condition is the same as Xiong et al~\cite{xiong2019biased}. 
Participants were shown a line chart for 0.5 to 1.6 seconds. The chart was then covered with a visual mask and after 0.5 seconds, the visual mask was replaced by an empty chart. The participants were asked to place a horizontal line on the empty chart using the cursor where they thought the average line position was. 

In the With-AI condition, we provided predictions from an AI model and controlled the AI-model's prediction time to assemble the real-world conditions when AI might be running on different devices and may have different response time. 

\paragraph{Temporal Delay.}
Specifically, we introduced a random temporal delay before the participant could see the AI recommendation. 

\paragraph{Procedure.}
The With-AI tests were conducted in the same way as the control, but after the mask was replaced by an empty chart, and followed by a delay of 0 to 5 seconds, the prediction by AI was displayed as a red horizontal line. The red horizontal line stayed on screen for 0.5 to 1.6 seconds before being replaced by an empty chart. Finally, the participant was asked to place their estimate for the true mean on the empty chart. Figure 2 illustrates the decision-making phase of the AI test after the visual mask is removed. In this specific case, the participant chose the perceived average to be lower than that of the AI despite seeing the AI prediction displayed.





\subsection{Visualizing AI Model's Insight}

\begin{figure}[tb]
 \centering 
 \includegraphics[width=0.95\columnwidth]{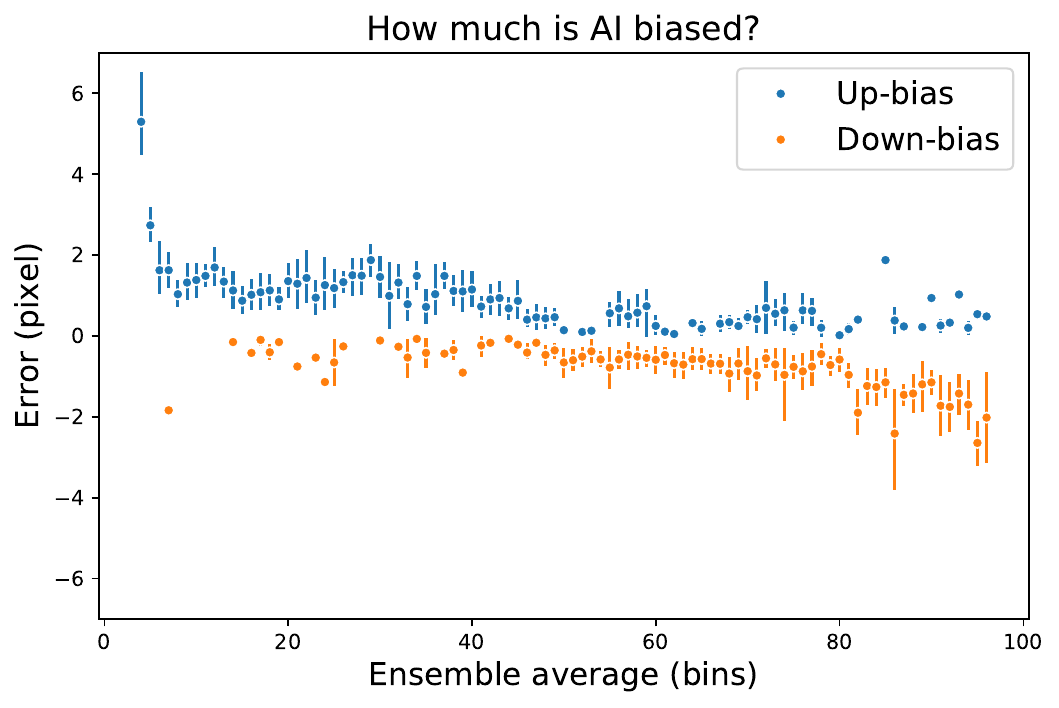}
 \vspace{-10pt}
 \caption{Results from Step 1 for evaluating AI's responses. On average AI's absolute error is considerably low with only a few pixels off  on a $100 \times 100$ $pixel^2$ graph.}
 \label{fig:aiResult}
\end{figure}

In our experiment, we used real AI responses and added some uncertainty to the AI responses before sending the prediction to our one-off AI visualization.  
\paragraph{Dataset and Model Implementation.} To train our convolutional neural network (CNN) model to answer ensemble average questions, we used our visualization program for $100\times100 pixel^2$ size, to generate 12,000 sample graphs with ground truths: 10,000 train, 1,000 validation, and 1,000 test. The CNN model  was VGG19~\cite{simonyan2014very} pretrained on ImageNet~\cite{5206848}. We send our visualizations to supervise VGG19 and used VGG19 for feature extraction and added a regression head to estimate the mean. The model was implemented in TensorFlow using the Keras deep learning API. 
The graphs shown on the screen in Xiong et al. had smaller vertical bar heights and occupied a limited screen size. We thus scaled our results to match the graphs in human testing. In general, AI's results were less than 1-pixel on the screen from the groud-truth. 

\paragraph{AI Model Performance.} The model was considerably accurate and was able to achieve a mean absolute error that is equivalent to 0.3 pixels off the true average, although errors increased as the mean line position approached the two ends (Figure~\ref{fig:aiResult}).

\paragraph{AI Model Shows No Downward Biases.} 
We examined whether or not the AI predictions tended to underestimate the true means. 
A Welch’s T-Test showed a $p$ = 0.8 for the null hypothesis of having a mean difference of 0 between the ground truths and AI predictions. This indicates that AI does not have the ``pull-down'' biases as humans do.

\paragraph{Addition of Noise to AI Assistance.}
In the empirical study, we added a random noise between $+/- 3\%$ of the chart range to the AI prediction when showing it as a red line to the display. This was to avoid participants' overreliance on AI and to see how far away the AI could stray from its prediction before the participant would no longer trust it.

\section{Results}

In this pilot study, we collected 300 data points for the ensemble average tasks, including response time and participants’ predictions, from six volunteers. 
In preparing the accuracy and task completion time for analysis, we computed participants' errors. 
We used the pair-wise t-test to examine the accuracy and task-completion time for the With-AI and Without-AI conditions. 
When the dependent variable was
binary (i.e., answer correct or wrong), we used a logistic
regression and reported the p-value from the Wald $\chi^2$ test.
When the p-value was less than 0.05, variable levels with
95\% confidence interval of odds ratios not overlapping were
considered significantly different. All error bars represent
$95\%$ confidence intervals.

\begin{figure}[tb]
 \centering 
 \includegraphics[width=0.95\columnwidth]{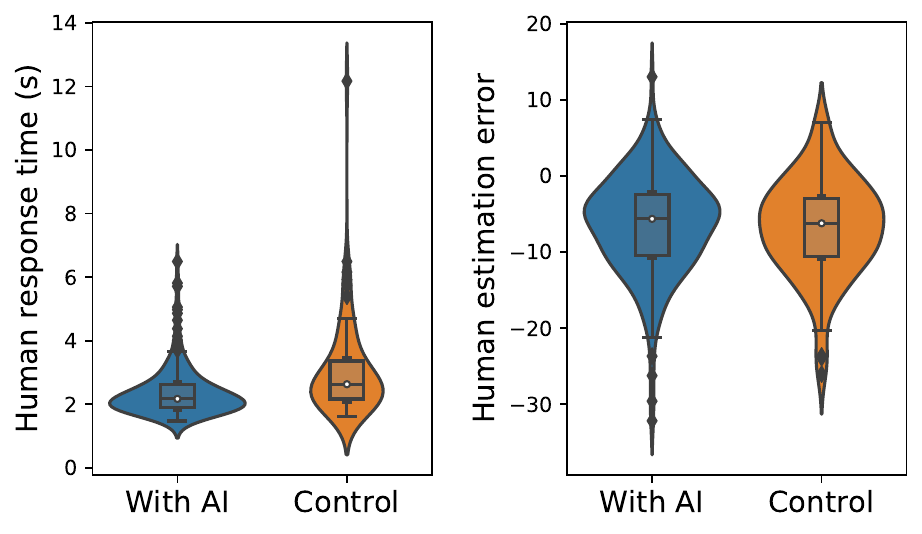}
 \vspace{-10pt}
 \caption{Comparison of results for the With-AI and Without-AI (Control) conditions. Left: response time; right: estimation error.}
 \label{fig:humanResult}
\end{figure}

\textbf{Participants were Faster with AI Assistance.}
The pilot study results supported the second half of our first hypothesis, i.e., there were observed improvements in response time for the AI-assisted estimates (Figure~\ref{fig:humanResult}, left plot). 
When we performed a 
t-test on the null hypothesis that the response times were the same between the control and AI-assisted tests, the result reported $p<0.0001$.
From this we conclude that the response times with the AI assistance were much faster (mean times: 3.05 seconds for the control group and 2.38 seconds for the With-AI group).
 

\textbf{Human Biases Persisted Regardless of AI Assistance. }
As shown in Figure~\ref{fig:humanResult} right plot, human estimation errors were mostly below zero and had a similar distribution with or without AI assistance. Specifically, $85.56\%$ of the Without AI estimates were underestimates, and $83.33\%$ of the With AI estimates were underestimates. The mean error was -6.14 for the Without-AI condition and -5.96 for the With-AI condition; the slight decrease in error was too small and not statistically significant despite a small sample size. This result supported our H2 (human pull-down biases stayed) but was inadequate to support the first half of H1, that human performance would improve with AI assistance.
Further, linear regressions on our collected data did not report significant p-values to define the relationship between AI prediction or AI pseudo-response time and response time or estimation error. 
H3 (AI-debias) was not supported since we did not observe improvement when AI was present.

\textbf{Pull-down Biases Existed for a Longer Delayed-Viewing Time.} In the experiment of Xiong et al.~\cite{xiong2019biased}, participants were given only 0.5 seconds for viewing graphs. In this study, participants were give a viewing time up to three times longer delay and the pull-down biases still existed. This may be suggestive that the human pull-down biases exist regardless of AI responses - observation time. It is unclear if humans ``herd'' with AI (H4 was not supported). 

%

\paragraph{Subjective Comments.}
Participants offered varying opinions on the prediction. Comments ranged from, ``\textit{I thought the red line did a good job, so I tried to match it closely}'', to ``\textit{I thought the red line was unhelpful and ignored it.}''  Again, the participants neither know that it was an AI that generated the prediction nor know that we added noise to the AI predictions (just 1-pixel away from ground-truth). Most participants did not even notice the temporal delay. 

\section{Discussion}


This section discusses the design knowledge we can glean from our pilot study on hypothesis testing and its ability to address our research questions in future experiments. 

\subsection{Encoding AI Insights}

We arrived at the following set of encoding aims for AI insights while considering the next steps from our observations. 

\begin{itemize}
    \item \textit{Aim 1. Represent AI insights using a distinct cue. } 
    It is crucial to present AI insights in a way that they can be differentiated from the original data. In this fashion, people can have better judgment choosing to use or ignore AI's inferences. In our situation, we should have informed humans the red-lines were AI predictions. 
    \item \textit{Aim 2. Proximity to task data.} Since task data is what draws viewer's attention, AI insights would be more accessible if overlaid on top of or in closer proximity to task data.         
    \item \textit{Aim 3. Avoid blocking the data view.} We envisioned that multi-layer representations~\cite{kirby1999visualizing} or other artistic rendering~\cite{lu2003illustrative} may be more suitable such that AI insights can be displayed in the background to inform people without blocking the data view. 
    \item \textit{Aim 4. Improve its ability to reveal global features.} Since AI insights are shown directly, global scene features can be revealed pre-attentively. We could utilize this in a subtle yet prominent way to produce visually dominant displays. However, they should not alter the global ensemble data perception except making it perceptually accurate. 
\end{itemize}

Traditional approaches to visualizing data lack the above considerations for encoding AI results.
It is vital to ensure that people viewing in the AI assistance environment can make independent decisions effectively alongside the data. 
Designers may also choose to make AI insights more prominent in conditions where AI's inferences are more accurate (e.g., elementary graphical perception tasks). 
Many computer vision solutions show AI results as spatial highlights. Line-based displays and heatmap-based approaches might occlude the task data and thus could be poor choices.

In future work, the number of participants and number of questions should be expanded. Other changes might also improve the ecological validity of the study, e.g., showing the ensemble and groundtruth in one chart without blinking the chart shortly before performing the task. 


\subsection{Balanced Controlled Study and Piratical Uses}
We plan to run conditions without the artificial $+/-3\%$ AI noises since we anticipate that AI models are reasonably accurate for ensemble perception, to avoid conditions that may not exist in real-world uses. 
Another interesting experiment is gradually increasing AI noises, and finding out the level of departure from the groundtruth when AI suggestion will be confidently rejected by the participants. 
The last issue we found with the experiment is the judgment for visual estimation on small graphs. The charts disappeared after each trial before the participants gave an answer. In conditions like this, participants had to estimate based on what they remember of the original data. Braun et al. had a more thorough treatment of these conditions 
~\cite{braun2023visual}. We plan to make the graph bigger and present on the screen in future experiments. 
We also plan to test harder tasks, such as correlation estimates from scatterplots. 



The overall goal of intelligent augmentation 
is also to enable the AI to reason for its decisions, so the human can justify it as part of their decision. 
However, none of our real-time decision making systems has provided justification. An interesting method of investigation is to show the participants their decision and response time for the previous test, before they can make a decision on the current test. 
Such an ``understand-then-respond'' procedure requires us to pay extra attention to the effectiveness of the initial visualization in the experiment. 
The performance of the experiment will be significantly undermined unless the participants can achieve a high level of understanding towards a visualization within a short period of time. 

\section{Conclusion}

The results of our study on line chart position estimation showed that the estimation assisted with AI prediction did lead to faster response time. The results, however, were not statistically significant to confirm an improvement in estimation accuracy or bias. 
While the increased viewing time was primarily to maintain consistency and allow our participants to perceive multiple things, error results on the control group extend the finding of Xiong et al.~\cite{xiong2019biased} that humans tend to underestimate the true average of a line chart. Further improvement is necessary on the assistance of AI to human visual perception tasks. 


\bibliographystyle{abbrv-doi}

\bibliography{ms}
\end{document}